\documentclass[english,twocolumn,showpacs,nofootinbib,superscriptaddress,showpacs]{revtex4}

\usepackage{latexsym}
\usepackage{graphicx}
\usepackage{epsfig,subfig}
\usepackage{epstopdf}
\usepackage[active]{srcltx}
\usepackage{amsmath}
\usepackage{amssymb}
\usepackage{hyperref}
\usepackage{cleveref}

\begin{document}

\title{Tunneling of Polymer Particles}
\author{A. Mart\'{\i}n-Ruiz}
\affiliation{Instituto de Ciencias Nucleares, Universidad Nacional Aut\'{o}noma de M\'{e}%
xico, 04510 M\'{e}xico, D.F., Mexico}
\email{alberto.martin@nucleares.unam.mx}

\author{E. Chan-L\'{o}pez}
\affiliation{Divisi\'{o}n Acad\'{e}mica de Ciencias B\'{a}sicas, Universidad Ju\'{a}rez Aut\'{o}noma de Tabasco, 86690 Cunduac\'{a}n, Tabasco, México.}

\author{A. Carbajal-Dom\'{i}nguez}
\affiliation{Divisi\'{o}n Acad\'{e}mica de Ciencias B\'{a}sicas, Universidad Ju\'{a}rez Aut\'{o}noma de Tabasco, 86690 Cunduac\'{a}n, Tabasco, México.}

\author{J. Bernal}
\affiliation{Divisi\'{o}n Acad\'{e}mica de Ciencias B\'{a}sicas, Universidad Ju\'{a}rez Aut\'{o}noma de Tabasco, 86690 Cunduac\'{a}n, Tabasco, México.}

\begin{abstract}
In this paper we study the tunneling using a background independent (polymer) quantization scheme. We show that at low energies, for the tunneling through a single potential barrier, the polymer transmission coefficient and the polymer tunneling time converge to its quantum-mechanical counterparts in a clear fashion. As the energy approaches the maximum these polymer quantities abruptly decrease to zero. We use the transfer matrix method to study the tunneling through a series of identical potential barriers. We obtain that the transmission coefficients  (polymer and quantum-mechanical)  behave qualitatively in a similar manner, as expected. Finally we show that the polymer tunneling time exhibits anomalous peaks compared with the standard result. Numerical results are also presented. 
\end{abstract}

\maketitle

\section{Introduction}

One of the most important goals of modern theoretical physics is to reconcile two of its cornerstones: General Relativity (GR) and Quantum Field Theory (QFT). Although both theories have been successful in explaining and predicting the observed phenomena with a high degree of accuracy, they come with their own set of deficiencies: ultraviolet divergences in QFT and singularities in GR. It is generally believed that a full Quantum Theory of Gravity (QTG) will solve these outstanding problems.

Quantum gravitational effects are expected to become relevant near the Planck scale, where spacetime itself is assumed to be quantized. The most popular approach to quantum gravity is String Theory (ST), where the fundamental particles are assumed to be extended objects rather than being point-like. A background independent quantization scheme that arose in Loop Quantum Gravity (LQG), the so called Polymer Quantization (PQ), has been used to explore mathematical and physical implications of theories such as quantum gravity \cite{Hossain2, Chacon}. The huge discrepancy between the Planck energy and the typical energy scales we are able to reach in our experiments make it virtually impossible to test these theories. However a possible route to test quantum gravitational effects is through deviations from the standard theory. 

Polymer quantization has attracted some attention in recent years in the fields dealing with the quantum gravitational effects in a physical system. In this framework, in Ref.\cite{Chacon} have been studied the quantum gravitational corrections to the standard thermodynamical quantities. On the other hand, in Ref.\cite{DIT} the author studies the quantum gravitational corrections to the temporal dynamics of a well-known quantum transient phenomena, the Diffraction in Time. Of course, such corrections depend on the polymer length scale. Based on these approaches, the purpose of this paper is to consider one of the simplest quantum-mechanical phenomena, the tunneling through a rectangular potential barrier, in order to explore if the polymer theory induces whether or not significant deviations from the well former quantum theory that could be detected in lab.

The remainder of the paper is organized as follows. In \cref{singleB}, the tunneling through a single rectangular potential barrier will be discussed. \Cref{Nbarrier} is devoted to the tunneling through a series of rectangular potential barriers. Finally, \cref{discussion} summarizes the results of this work and draws conclusions.

\section{Tunneling through a single potential barrier} \label{singleB}

Let us consider a polymer particle of mass $m$ with energy $E$ which incides upon a rectangular potential barrier of height $ U _{0} > E $ and width $ L $. 

To address this problem, we restrict the dynamics to an equispaced lattice $ \gamma \left( \lambda \right) = \left\lbrace \alpha \lambda \vert \alpha \in\mathbb{Z} \right\rbrace $. The spectrum of the position operator $ \left\lbrace x _{\mu} = \mu \lambda \right\rbrace $ consists of a countable selection of points from the real line. Here $ \lambda $ is regarded as a fundamental length scale of the polymer theory. In \cref{PQM} we present a brief review of Polymer Quantum Mechanics. In this framework the width $L$ is restricted to be multiple of the fundamental length, i.e. $L = n \lambda $, with $ n \in \mathbb{Z} ^{+} $.

[Hereafter we use the following dimensionless quantities for position, momentum, energy and time
\begin{equation}
\mu \equiv \frac{x _{\mu}}{\lambda} \;\;  , \;\; \rho \equiv \frac{p \lambda}{\hbar} \;\; , \;\; \varepsilon \equiv \frac{m \lambda ^{2} E}{\hbar ^{2}} \;\;  , \;\;  \tau = \frac{\hbar t}{m \lambda ^{2}} \label{dimensionless}
\end{equation}
respectively.]

We can begin the analysis of this set‐up by using the time-independent polymer Schr\"{o}dinger equation in coordinate representation
\begin{equation}
\psi _{\mu + 1} + \psi _{\mu - 1} = 2 \left[ 1 - \left( \varepsilon - \upsilon \right) \right] \psi _{\mu}, \label{polymer schrodinger eq}
\end{equation}
where 
\begin{equation}
\upsilon \left( \mu \right) = \upsilon _{0} \Theta \left( \mu \right) \Theta \left( n - \mu \right) \label{potential}
\end{equation}
is the barrier potential with height $ \upsilon _{0} > \varepsilon $. $\Theta \left( x \right) $ is the Heaviside step function.

Denoting by $ I $, $ II $ and $ III $ the regions $ \mu \leq 0 $, $ \mu \in \left[ 0 , n \right]  $ and $ \mu \geq n $, respectively, the wave functions for these regions are
\begin{eqnarray}
\psi ^{I} _{\mu} &=& e ^{i \rho \mu } + R e ^{-i \rho \mu}, \\ \nonumber
\psi ^{II} _{\mu} &=& A e ^{-\kappa \mu} + B e ^{\kappa \mu}, \\ \nonumber
\psi ^{III} _{\mu} &=& T e ^{i \rho \mu}, \label{waves region}
\end{eqnarray}
where $ R $ and $ T $ are the reflected and transmitted amplitudes, respectively. Here $ \rho $ and $ \kappa $ satisfy the polymer dispersion relations
\begin{eqnarray}
\varepsilon &=& 1 - \cos \rho , \\ \nonumber
\varepsilon - \upsilon _{0} &=& 1 - \cosh \kappa , \label{dispersion rel}
\end{eqnarray}
considering a fixed value of $ \varepsilon < \upsilon _{0} $. Note that the free energy spectrum is bounded from above ($\varepsilon _{\max} = 2$), and the bound depends on the length scale $\lambda$.

\begin{figure}[tbp]
\vspace{0.5 cm}
\par
\begin{center}
\includegraphics[scale=0.5, natwidth=640, natheight=480]{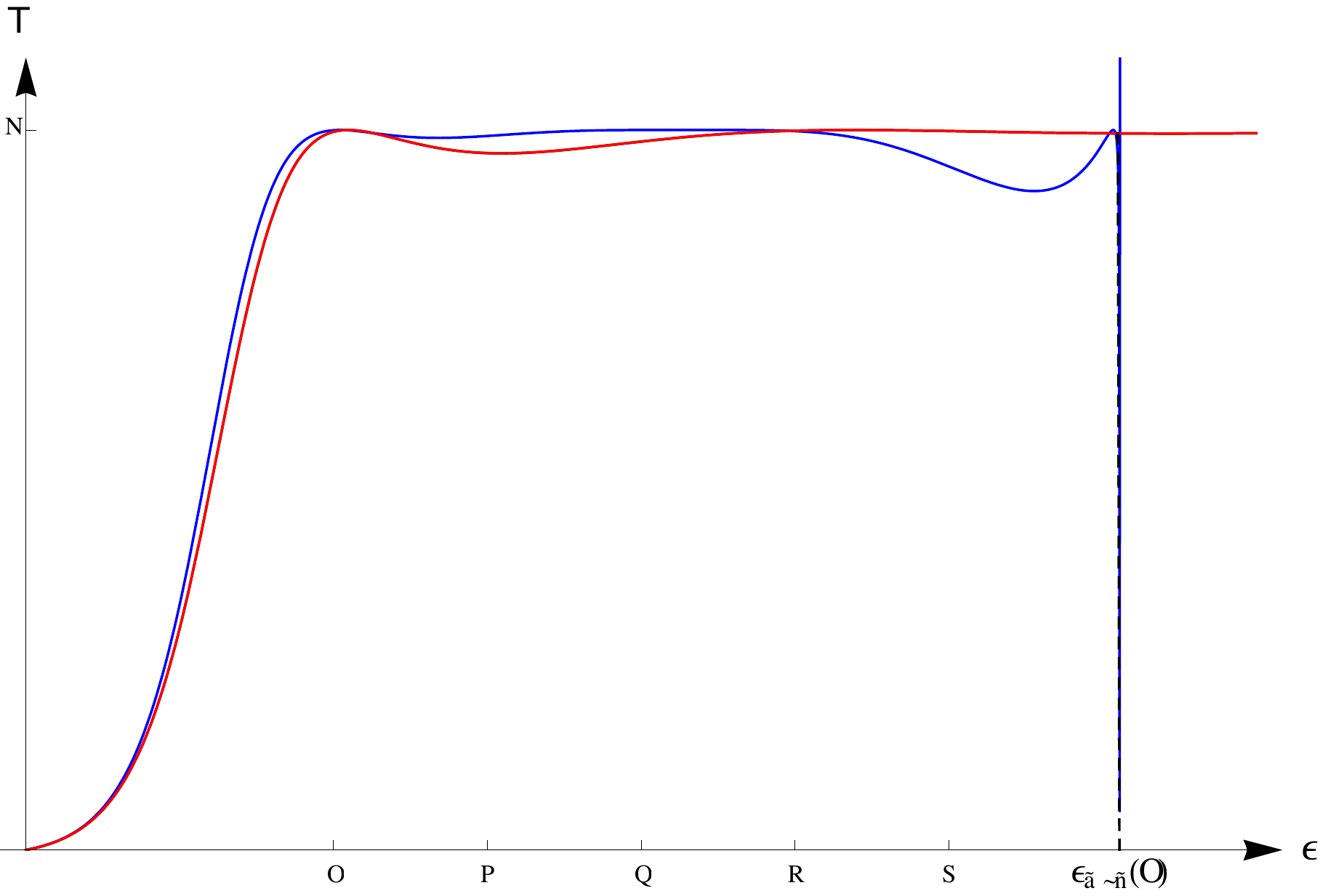}
\end{center}
\caption{\small Plots of the quantum-mechanical (black line) and the polymer transmission coefficients (the red line for $n=100$ and the blue line for $n=2$) as a function of the ratio $\epsilon = \frac{E}{U _{0}}$. } \label{TC 1B}
\end{figure}

As usual, for finding the amplitudes we must apply the appropriate boundary conditions on the lattice. The continuity of the wave functions is needed as in the standard quantum theory, but the continuity of spatial derivative must be replaced by its discretized version. Then we have the conditions
\begin{eqnarray}
\psi ^{I} _{0} &=& \psi ^{II} _{0} , \\ \nonumber
\psi ^{II} _{n} &=& \psi ^{III} _{n} , \\ \nonumber
\psi ^{I} _{0+1} - \psi ^{I} _{0-1} &=& \psi ^{II} _{0+1} - \psi ^{II} _{0-1} , \\ \nonumber
\psi ^{II} _{n+1} - \psi ^{II} _{n-1} &=& \psi ^{III} _{n+1} - \psi ^{III} _{n-1} , \label{boundary conditions}
\end{eqnarray}
which produce a set of four simultaneous equations. The solution for the transmission amplitude is
\begin{equation}
T = \frac{ e ^{ -i \rho n} }{ \cosh \left( \kappa n \right) + i \frac{ \xi ^{2} - \sigma ^ {2} }{2 \xi \sigma} \sinh \left( \kappa n \right) } , \label{transmission amplitude}
\end{equation}
where $ \sigma = \sin \rho $ and $ \xi = \sinh \kappa $. The corresponding polymer transmission coefficient is then $\mathcal{T} = \vert T \vert ^{2} $. For comparison with the standard result, in \cref{TC 1B} we plot the quantum-mechanical and the polymer transmission coefficients for an electron incident upon a rectangular barrier of height $U _{0} =10$eV and thickness $L = 1.8 \times 10 ^{-10}$m. For the polymer result we plot two cases, $n = 2$ (blue line) and $n= 100$ (red line). This rectangular barrier is an idealization of the barrier encountered by an electron that is scattering from a negatively ionized gas atom in the ``plasma" of a gas discharge tube. The actual barrier is not rectangular, of course, but it is about the height and thickness quoted \cite{Eisberg}.

We observe that at low energies ($n \gg 1$), the polymer and standard cases behave qualitatively in a similar manner, as expected. Indeed at second order of approximation, the polymer transmission coefficient becomes
\begin{equation}
\mathcal{T} \sim \mathcal{T} ^{QM} - \frac{2}{3} \frac{\alpha}{n ^{2}} \frac{ \epsilon ^{2} - \left( 1-\epsilon \right) ^{2} }{\epsilon \left( 1 - \epsilon \right) } \sinh ^{2} \left( \sqrt{\alpha \left( 1- \epsilon \right) } \right) , \label{low energy TC}
\end{equation}
where $\mathcal{T} ^{QM}$ is the standard quantum mechanical result, $ \alpha \equiv \frac{2 m L ^{2} U _{0}}{\hbar ^{2}} $ and $\epsilon \equiv \frac{E}{U _{0}}$. For the case $n=100$ considered in \cref{TC 1B}, the polymer correction is of the order of $10 ^{-4}$, which is extremely small to be detected in lab. Moreover taking $\lambda$ in the order of the Planck length ($l _{p} = 1.6 \times 10 ^{-35}$m), such deviation is extremely small ($\mathcal{T} - \mathcal{T} ^{QM} \approx l _{p} ^{2}$).

On the other hand, at high energies ($n$ small) the polymer effects become important, as we can see in \cref{TC 1B} (blue line, $n=2$). The most astonishing result is that the polymer transmission coefficient decreases abruptly to zero when the energy approaches the maximum, i.e. at $\epsilon _{\max} \left( n\right)  = \frac{4 n ^{2}}{\alpha}$, while the quantum-mechanical result remains at one. Compared with the typical energies we are able to reach in our experiments, $\epsilon _{\max} $ is too high to hope to be able to test such effect.

As any deviation from the standard theory is, at least in principle, experimentally testable, now we study the time-delay caused by tunneling. The analysis of tunneling time is complicated because time plays an unusual and subtle role in quantum theory. Unlike the position (represented by a hermitian operator), time is represented by a $c$-number. Consequently, although the time-energy uncertainty relation is similar in appearance to the familiar position-momentum  uncertainty relation, its origin and interpretation is quite different. In this work we consider the usual procedure introduced by Salecker and Wigner \cite{wigner} for calculating the tunneling time. In Ref.\cite{davies}, the author presents a complete review of the Salecker-Wigner procedure. Other possible ways of definig the tunneling time are reviewed in Ref.\cite{olkhovsky, razavy}.

\begin{figure}[tbp]
\vspace{0.5 cm}
\par
\begin{center}
\includegraphics[scale=0.5, natwidth=640, natheight=480]{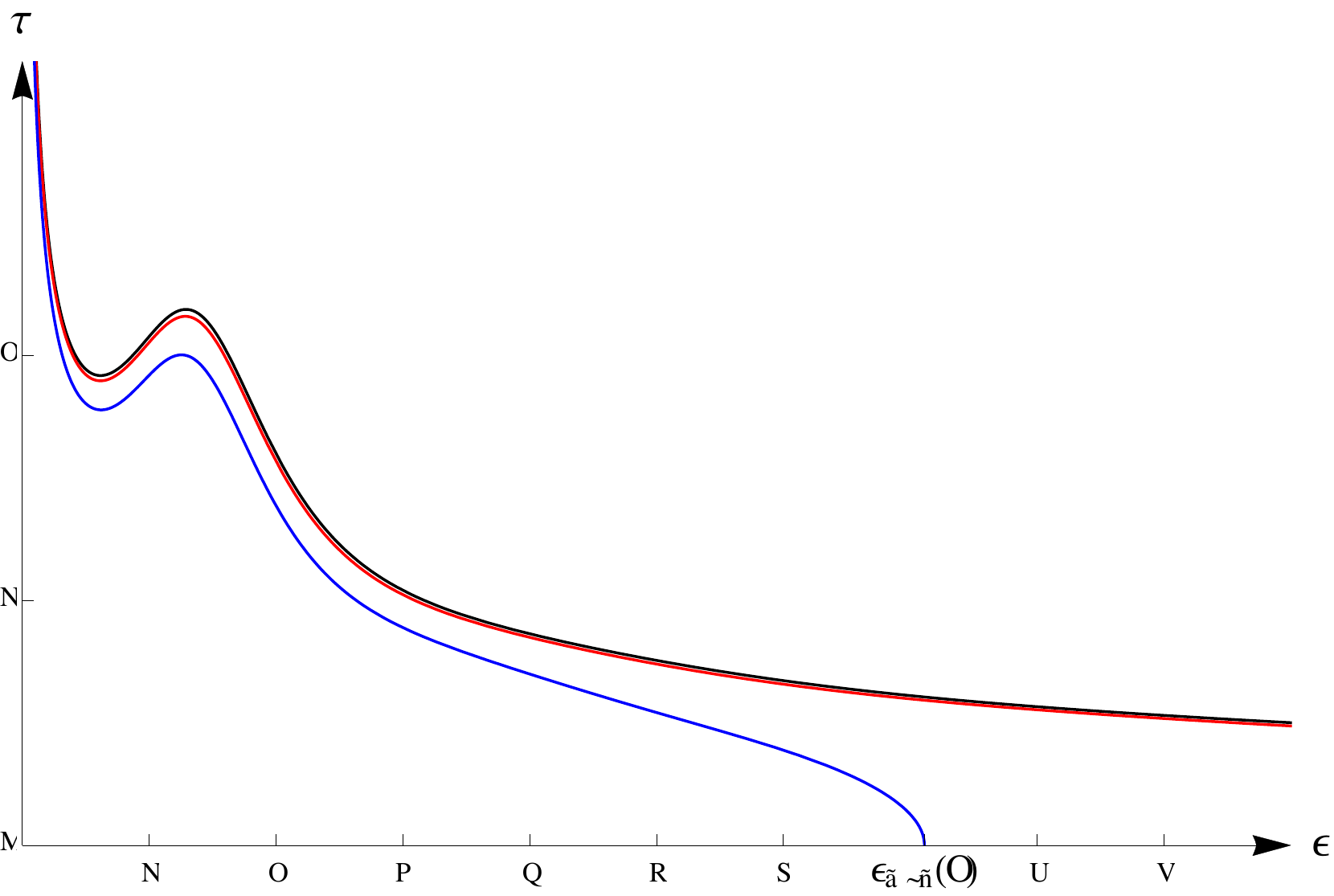}\qquad
\end{center}
\caption{\small Plots of the quantum-mechanical (black line) and the polymer tunneling time (the red line for $n=100$ and the blue line for $n=2$) as a function of the ratio $\epsilon = \frac{E}{U _{0}}$. } \label{TT 1B}
\end{figure}

The phase difference of the wave function between the region $I$ and $III$ is
\begin{equation}
\delta \left( \epsilon \right) = - \arctan \left[ \frac{\xi ^{2} - \sigma ^{2}}{2 \xi \sigma} \tanh \left( \kappa n \right) \right], \label{phase change}
\end{equation}
where we have used the transmission amplitude (\ref{transmission amplitude}). The tunneling time is defined as $ \frac{d \delta \left( \epsilon \right) }{d \epsilon}$. By differentiating (\ref{phase change}) we find that the expectation value of the polymer tunneling time is
\begin{equation}
\tau = \frac{2n \sigma \left( \xi ^{2} - \sigma ^{2} \right) + \left( \xi ^{2} + \sigma ^{2} \right) \left( \frac{\xi}{\sigma} \cos \rho + \frac{\sigma}{\xi} \cosh \kappa \right) \sinh \left( 2 \kappa n \right) }{\left( \xi ^{2} + \sigma ^{2} \right) ^{2} \cosh ^{2} \left( \kappa n \right) - \left( \xi ^{2} - \sigma ^{2} \right) ^{2} } \label{time}
\end{equation}
Of course, at low energies $\tau$ reduces to the quantum tunneling time. In \cref{TT 1B} we plot both, the polymer and the quantum-mechanical tunneling time for the system considered before. We can see that at low energies both cases behave qualitatively in a similar manner. As we increase energy, the polymer tunneling time starts to deviate from its quantum-mechanical counterpart. Also we observe that $\tau$ decreases to zero when the energy approaches the maximum, while the quantum-mechanical result remains in a finite value.
So far we have seen that the polymer effects become important at high energies, however it would be interesting if the polymer effects could be amplified at low energies. To this end, in the next section we will consider the tunneling through a series of identical potential barriers.

\section{Tunneling through a series of identical potential barriers} \label{Nbarrier}

Let us assume that there are $N$ rectangular barriers each of height $U _{0}$ and width $L = n \lambda$, and the distance between the two barriers is $l = m \lambda$, with $n,m \in \mathbb{Z} ^{+}$. For solving the problem we first find the transfer matrix $\textbf{T} ^{(1)}$ for a single barrier. The polymer wave function for the three regions are
\begin{eqnarray}
\psi ^{I} _{\mu} &=& A _{I} e ^{i \rho \mu } + B _{I} e ^{-i \rho \mu}, \\ \nonumber
\psi ^{II} _{\mu} &=& A _{II} e ^{-\kappa \mu} + B _{II} e ^{\kappa \mu}, \\ \nonumber
\psi ^{III} _{\mu} &=& A _{III} e ^{i \rho \mu } + B _{III} e ^{-i \rho \mu}, \label{waves region 2}
\end{eqnarray}
and the transfer matrix $\textbf{T} ^{(1)}$ is defined by
\begin{equation}
\left[ \begin{array}{c} A _{III} \\ B _{III} \end{array} \right] = \textbf{T} ^{(1)} \times \left[ \begin{array}{c} A _{I} \\ B _{I} \end{array} \right]. \label{transfer}
\end{equation}
With the help of boundary conditions (\ref{boundary conditions}), we obtain the elements of the transfer matrix as
\begin{eqnarray}
T _{11} ^{(1)} &=& T _{22} ^{(1)\ast} = \left[ \cosh \left( \kappa n \right) + i \frac{\sigma ^{2} - \xi ^{2}}{2 \sigma \xi} \sinh \left( \kappa n \right)  \right] e ^{-i \rho n} , \\ \nonumber
T _{12} ^{(1)} &=& T _{21} ^{(1)\ast} = -i \frac{\sigma ^{2} + \xi ^{2}}{2 \sigma \xi} \sinh \left( \kappa n \right) e ^{-i \rho n}, \label{Telements}
\end{eqnarray}
and also that $ \mathrm{det} \; \textbf{T} ^{(1)}  = 1 $.

Now, let us generalize the problem to multiple potential barriers. The transfer matrix $\textbf{T} ^{(N)}$ which relates the amplitudes of the incoming and outgoing waves in the $N$ barrier system can be defined through
\begin{equation}
\left[ \begin{array}{c} A _{N} \\ B _{N} \end{array} \right] = \textbf{T} ^{(N)} \times \left[ \begin{array}{c} A _{I} \\ B _{I} \end{array} \right] . \label{Ntransfer}
\end{equation}
The transfer matrix method discussed before can be extended and applied to the $N$ barrier system. We obtain that the transfer matrix can be expressed as $\textbf{T} ^{(N)} = \left( \textbf{F} ^{\ast} \right) ^{N} \textbf{G} ^{N}$, where 
\begin{equation}
\textbf{F} = \begin{bmatrix} e ^{i \rho \left( n+m -1\right) } & 0 \\ 0 & e ^{-i \rho \left( n+m -1\right) } \end{bmatrix} \label{Fmatrix}
\end{equation}
and $\textbf{G}  = \textbf{T} ^{(1)} \textbf{F} $. With the help of eq.(\ref{Telements}) and the diagonal form of $\textbf{F}$ we find that the matrix elements are
\begin{eqnarray}
\textbf{T} ^{(N)} _{11} &=& \textbf{T} ^{(N)\ast} _{22} = F _{22} ^{N} \; \left[ G _{11} \; U _{N-1}  - U _{N-2} \right] , \\ \nonumber
\textbf{T} ^{(N)} _{12} &=& \textbf{T} ^{(N)\ast} _{21} = F _{22} ^{N} \; G _{12} \; U _{N-1} , \label{TNcomponents}
\end{eqnarray}
where $G _{11} = T _{11} ^{(1)} \; F_{11} $, $G _{12} = T _{12} ^{(1)} \; F_{22} $,
\begin{eqnarray}
U _{N-1} \left( g\right)  &=& \frac{g _{+} ^{N} - g _{-} ^{N} }{g _{+} - g _{-} }, \\ \nonumber
g _{\pm} &=& g \pm \sqrt{g ^{2} - 1} , \label{definitions}
\end{eqnarray}
and $g = \mathrm{Re} \left[ G _{11} \right] $.

Depending on whether $\vert g \vert < 1$ or $\vert g \vert > 1$, we can rewrite $g _{+}$ and $g _{-}$ in the following ways: if $\vert g \vert < 1$ then
\begin{equation}
g _{+} = g _{-} ^{-1} = e ^{i \theta}  \; , \; \cos \theta = g , \label{g>1}
\end{equation}
and $U _{N-1} \left( \cos \theta \right)$ is the Chebyshev polynomial of the second kind \cite{Gradshtein}. If $\vert g \vert > 1$, then
\begin{equation}
g _{+} = g _{-} ^{-1} = e ^{ \varphi}  \; , \; \cosh \varphi = g , \label{g<1}
\end{equation}
but
\begin{equation}
U _{N-1} \left( \cosh \varphi \right) = \frac{\sinh \left( N \varphi \right) }{\sinh \varphi} . \label{Ug<1}
\end{equation}
The transmission amplitude $T _{N}$ across $N$ barriers can be obtained as $\frac{A _{N}}{A _{1}}$, but also imposing not reflected polymer particles beyond the right end of the multibarrier system (i.e. $B _{N} = 0$). Using this fact together with \cref{Ntransfer}, the transmission coefficient $ \mathcal{T} _{N}$ can be obtained as
\begin{equation}
\mathcal{T} _{N} = \frac{1}{ \vert T _{11} ^{(N)} \vert ^{2}} = \frac{1}{1 + \vert \textbf{T} ^{(1)} _{12} \vert ^{2} \vert U _{N-1} \vert ^{2} } . \label{TCNbarrier}
\end{equation}

\begin{figure}[tbp]
\vspace{0.5 cm}
\par
\begin{center}
\includegraphics[scale=0.5, natwidth=640, natheight=480]{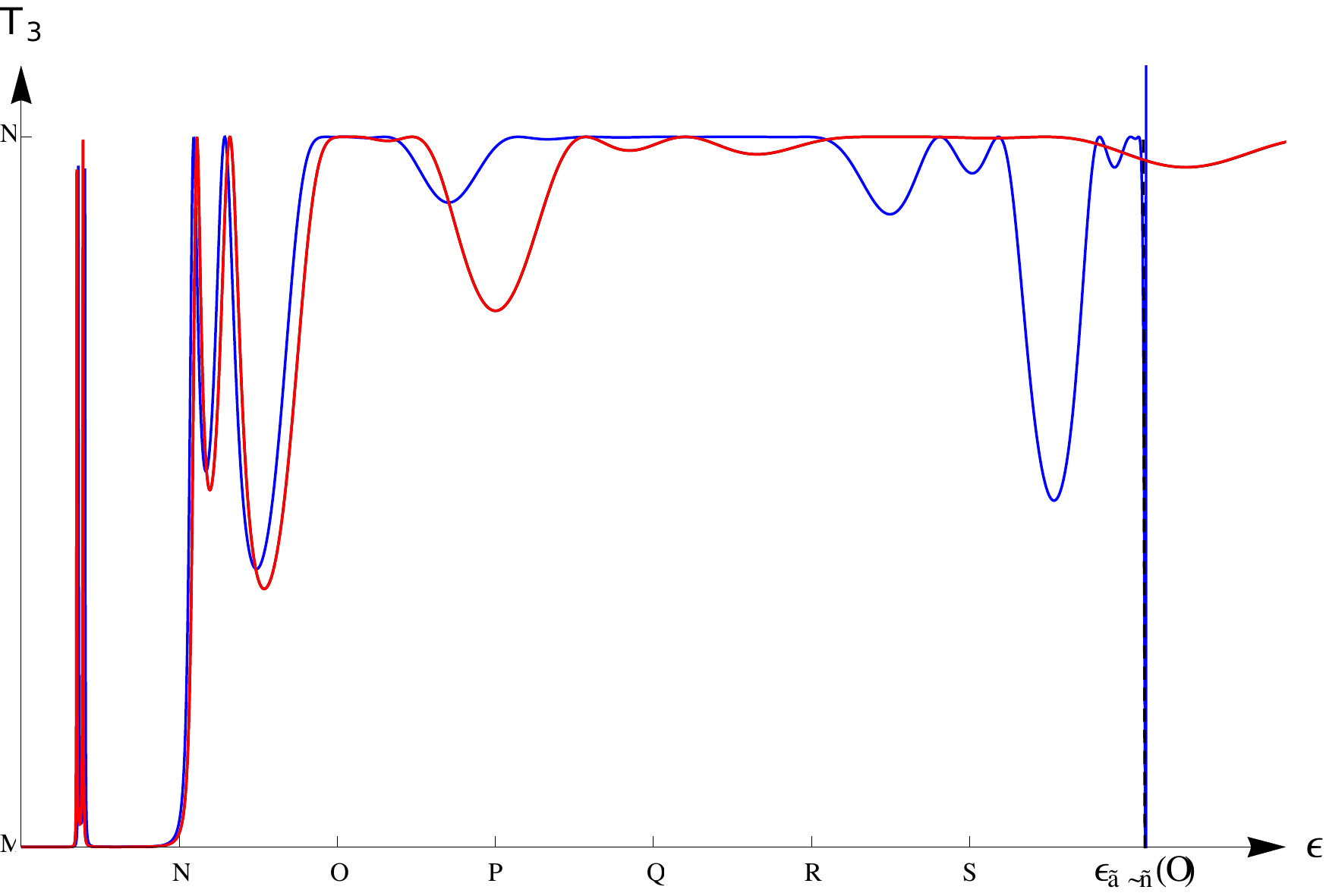}\qquad
\end{center}
\caption{\small Plots of the quantum-mechanical (black line) and the polymer transmission coefficients (the red line for $n=100$ and the blue line for $n=2$) for a series of three potential barriers as a function of the ratio $\epsilon = \frac{E}{U _{0}}$.} \label{TC 3B}
\end{figure}

In \cref{TC 3B} we present numerical results for the transmission coefficient for an electron incident upon a series of three rectangular barriers of height $U _{0} = 10$eV, thickness $L = 1.8 \times 10 ^{-10}$m and the distance between barriers $l=L$. As in the previous section, it is clear that at low energies the quantum-mechanical and the polymer transmission coefficients behave qualitatively in a similar manner. Also we observe that the polymer result decreases abruptly to zero near the maximum energy.

\begin{figure}[tbp]
\vspace{0.5 cm}
\par
\begin{center}
\includegraphics[scale=0.5, natwidth=640, natheight=480]{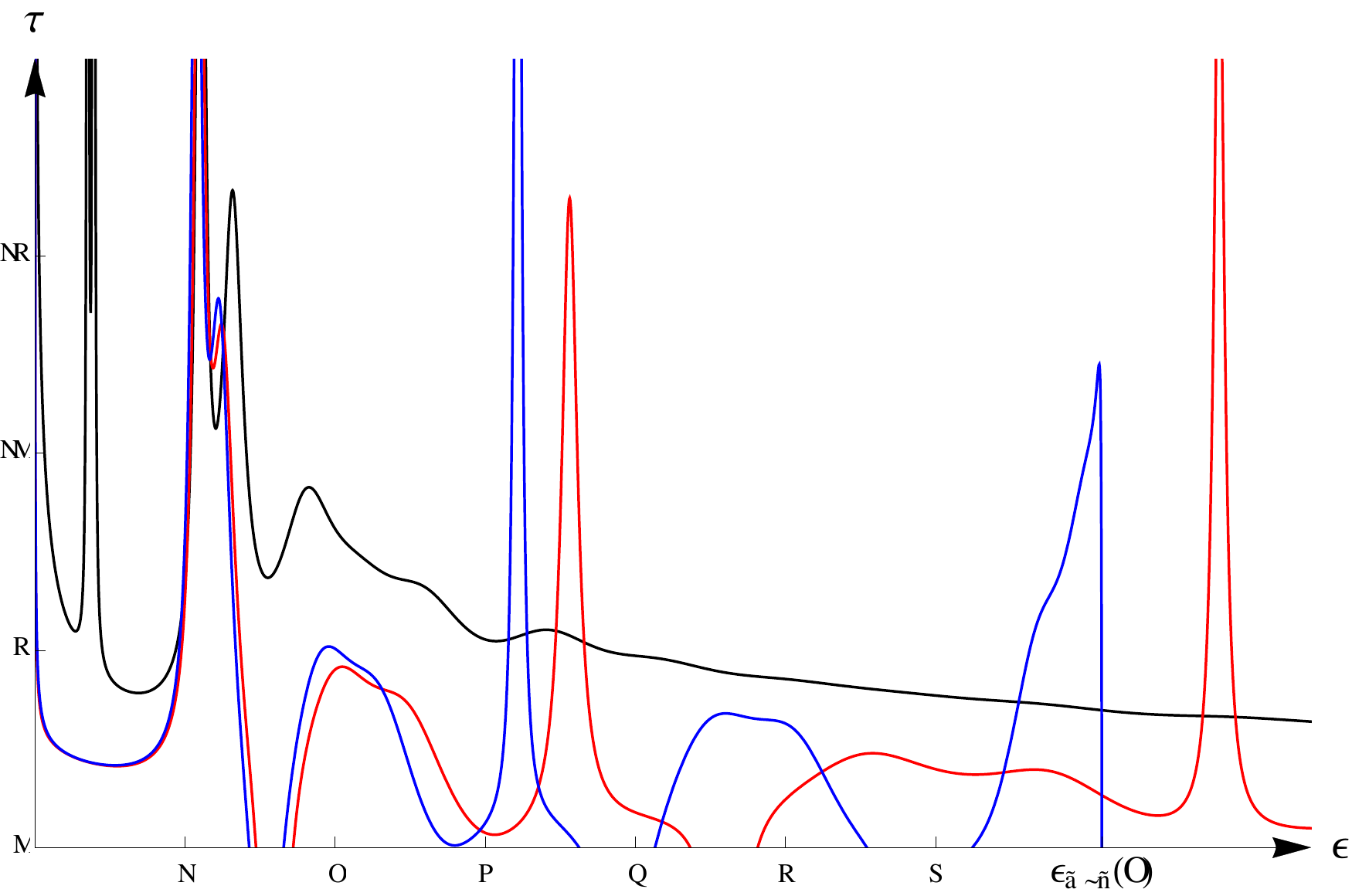}\qquad
\end{center}
\caption{\small Plots of the quantum-mechanical (black line) and the polymer tunneling time (the red line for $n=100$ and the blue line for $n=2$) for a series of three potential barriers as a function of the ratio $\epsilon = \frac{E}{U _{0}}$. } \label{TT 3B}
\end{figure}

The tunneling time can be analysed as in the previous section. The phase difference between the incident wave function at $\mu = 0$ and the transmited wave function at $\mu =N \left( n + m -1 \right) $ is
\begin{equation}
\delta \left( \epsilon \right) = - \arctan \left( \frac{\mathrm{Im} \left[ G _{11} \right]}{\mathrm{Re} \left[ G _{11} \right] - \frac{U _{N-2}}{U _{N-1}}} \right), \label{Nphase}
\end{equation}
which reduces to (\ref{phase change}) in the appropriate limit. In \cref{TT 3B} we superimpose the quantum-mechanical and the polymer tunneling time for an electron incident upon a series of three rectangular barriers. We observe that the polymer tunneling time is smaller than its quantum-mechanical counterpart, even at low energies. As before, it is clear that the polymer result decreases to zero as the energy approaches the maximum. An interesting finding is that (\ref{TCNbarrier}) exhibits anomalous peaks absent in the standard result. So far we have found that, for the system we have considered, the only significant difference between both theories is not through the transmission probability density, but through the tunneling time. 

\section{Discussion} \label{discussion}

The implementation of a minimal length scale in quantum theory constitutes a fundamental bound below which position can not be defined. It has been suggested that the position-momentum uncertainty relation should be modified to take into account the effects of spatial “grainy” structure. The implementation of such ideas in polymer quantum mechanics is a difficult task because the momentum operator is not directly realized as in Schr\"{o}dinger quantum mechanics. Some phenomenological aspects of effective QTG candidates have been introduced in quantum mechanics through deformation of the algebraic structure of ordinary quantum mechanics. For instance the GUP and non-commutative geometry are the most well known deformations that impose the ultraviolet and infrared cutoffs for the physical systems \cite{gorji}.

Assuming that the position-momentum Heisenberg uncertainty relation remains unchanged, the lower bound to the resolution of distances $\left( \Delta x \right) _{\min}  \sim \lambda$ introduces an upper bound to the resolution of momentum $\left( \Delta p \right) _{\max} \sim \hbar \lambda ^{-1}$, which in turn induces a minimal temporal window in the time-energy uncertainty relation given by $\left( \Delta t \right) _{\min} \sim \frac{m \lambda ^{2}}{\hbar}$. The small uncertainty in time implies a large uncertainty in energy $\left( \Delta E \right) _{\max} \sim \frac{\hbar ^{2}}{m \lambda ^{2}} $. The implementation of both, the position-momentum and the time-energy modified uncertainty relations, could play an important role in other branches of physics. In theories where there is no cutoff built-in, all states are expected to contribute to amplitudes with equal strength and consequently lead to UV infinities. A theory which naturally provides the adequate cutoffs built-in could shed light on the route for curing such UV divergences.

In summary, in this paper we have considered one of the simplest quantum-mechanical phenomena, the tunneling through a potential barrier. The aim of this work is to explore quantum gravitational corrections to the transmission coefficient and the tunneling time. In \cref{singleB} we find that at low energies, the polymer and the quantum-mechanical results are similar, but in the high energy regime the polymer effects take place. The most important result in this case is that the polymer quantities abruptly decreases to zero as the energy approaches the maximum. Of course, measuring this effect is now not yet feasible. In \cref{Nbarrier} we consider a series of $N$ identical potential barriers. Regarding the transmission coefficient, we observe basically the same behaviour in both cases, except as the energy approaches the maximum. The remarkable finding is that the polymer tunneling time is smaller than its quantum-mechanical counterpart, and also it exhibits anomalous peaks absent in the standard result.

\appendix

\section{Polymer Quantum Mechanics} \label{PQM} 

In the Schr\"{o}dinger representation of quantum mechanics the Hilbert space is $\textit{H} = L ^{2} \left(  \mathbb{R}, dx \right)  $ with Lebesgue measure $dx$. The central difference between the standard and polymer quantization is the choice of Hilbert space \cite{Hossain, Seahra}. In loop or polymer representation, the kinematical Hilbert space $\textit{H} _{poly}$ is the Cauchy completion of the set of linear combination of some basis states $ \left\lbrace \left\vert x_{\mu }\right\rangle \right\rbrace  $, with inner product
\begin{equation}
\left\langle x_{\mu} \vert x_{\nu} \right\rangle = \lim _{T \rightarrow \infty} \frac{1}{2T} 
\int _{-T} ^{T} dp e^{-i \frac{p}{\hbar} \left( x_{\mu} - x_{\nu} \right)} = \delta _{\mu 
\nu}, \label{inner}
\end{equation}
where $ \delta _{\mu \nu} $ is the Kronecker delta, instead of Dirac delta as in Schr\"{o}dinger representation, then we say that the orthonormal basis is discrete. Plane waves are normalizable in this inner product. The kinematical Hilbert space can be written as $ \textit{H} _{poly} = L^{2} \left(  \mathbb{R} _{d}, d \mu _{d} \right) $, with $  d \mu _{d} $ corresponding Haar measure, and $ \mathbb{R} _{d} $ the real line endowed with the discrete topology \cite{Chacon}.

The state of a polymer system can be expressed as
\begin{equation}
\left\vert \Psi \right\rangle = \sum _{\mu} \Psi \left( x_{\mu} \right)  \left\vert x_{\mu 
}\right\rangle .
\end{equation}
Here, $ \left\vert x_{\mu }\right\rangle $ are eigenstates of the position operator
\begin{equation}
\widehat{x} \left\vert x_{\mu }\right\rangle = x_{\mu } \left\vert x_{\mu }\right\rangle ,
\end{equation}
and the $ \Psi \left( x_{\mu} \right) $ are expansion coefficients. Note that the spectrum of the position operator $ \left\lbrace x _{\mu} \right\rbrace  $ consists of a countable selection of points from the real line $ \mathbb{R} $, which is analogous to the graph covering $3-$manifolds in LQG.

The central feature here is that the momentum operator $\widehat{p}$ is not realized directly as in Schr\"{o}dinger quantum mechanics because of built-in notion of discreteness, but arise indirectly through translation operator $\widehat{U} _{\lambda} \equiv e^{i \frac{\widehat{p} \lambda}{\hbar}} $ \cite{Hossain2}. Hence, for the representation of the Heisenberg-Weyl algebra we choose the position operator $ \widehat{x} $ and the translation operator $ \widehat{U} _{\lambda} $ instead of the momentum operator. The action of the translation operator on position eigenstates is
\begin{equation}
\widehat{U} _{\lambda}  \left\vert x_{\mu }\right\rangle = \left\vert x_{\mu } - \lambda 
\right\rangle ;
\end{equation}
that is, $ \widehat{U} _{\lambda} $ converts a position eigenstate with eigenvalue $ x_{\mu} $ into an eigenstate with eigenvalue $ x _{\mu} -\lambda $. These operators definitions give the basic commutator $ [ \widehat{x} , \widehat{U} _{\lambda} ] = -\lambda  \widehat{U} _{\lambda} $, and $ \widehat{U} _{\lambda} $ defines a one-parameter family of unitary operators on $ \textit{H} _{poly} $, where its adjoint is given by $ \widehat{U} ^{\dagger} _{\lambda} = \widehat{U} _{- \lambda}  $. Mathematically, polymer and Schr\"{o}dinger quantizations are inequivalent because $ \widehat{U} _{\lambda} $ is discontinues with respect to $ \lambda $ given that $ \left\vert x_{\mu }\right\rangle $ and $ \left\vert x_{\mu } - \lambda \right\rangle $ are always orthogonal, no matter how small is $ \lambda $ \cite{Corichi}.

However, inspired by the techniques used in Lattice Gauge Theories and LQG, by introducing a fixed length scale $ \lambda $ it is possible to define an effective momentum operator as follows
\begin{equation}
\widehat{p} _{\lambda} = \frac{\hbar}{2 i \lambda} \left( \widehat{U} _{\lambda} - \widehat{U} ^{\dagger} _{\lambda} \right) ,
\end{equation}
which corresponds to the approximation $ p \lambda \ll \hbar $.

In $ L^{2} \left(  \mathbb{R}, dx\right)  $, the $ \lambda \rightarrow 0 $ limit would give the usual momentum and momentum-squared operators $ -i \hbar \partial _{x} $ and $ - \hbar^{2} \partial ^{2} _{x} $ \cite{Hossain2}. In $ \textit{H} _{poly} = L^{2} \left(  \mathbb{R} _{d}, d \mu _{d} \right) $ this limit does not exist because $ \lambda $ is regarded as a fundamental length scale \cite{DIT}. This is analogous to the quantum-classical transition through $ \hbar \rightarrow 0 $ limit, where $ \hbar $ is a non-zero fundamental constant of quantum theory \cite{Martin}.

In order to study the dynamics of a physical system, we may proceed as in the standard case, with the dynamics determined by the Schr\"{o}dinger equation, i.e. $ i \hbar \partial _{t}  \left\vert \Psi \right\rangle = \widehat{H} _{\lambda} \left\vert \Psi \right\rangle $,  whose stationary solution $ \left\vert \Psi \right\rangle = e ^{-i \frac{Et}{\hbar}} \left\vert \psi \right\rangle $ are constructed from the energy eigenstates of the Hamiltonian operator \cite{Ashtekar} :
\begin{equation}
 \widehat{H} _{\lambda} = \frac{\hbar ^{2}}{2m \lambda ^{2}} \left( 2 - 
 \widehat{U} _{2\lambda} - \widehat{U} ^{\dagger} _{2 \lambda} \right)  + 
 \widehat{V} \left(  \widehat{x} \right) , \label{hamiltonian}
\end{equation}
where the potential term is arbitrary but assumed to be regular so that $ \widehat{V} $ can be defined pointwise multiplication, $ \left\langle x_{\mu} \left\vert \widehat{V} \right\vert \psi \right\rangle = V \left( x _{\mu} \right) \left\langle x_{\mu} \vert \psi \right\rangle  $.

The dynamics generated by (\ref{hamiltonian}) decomposes the polymer Hilbert space $ \textit{H} _{poly} $, into an infinite superselected finite-dimensional subspaces, each with support on a regular lattice $ \gamma = \gamma \left( \lambda , x _{0} \right)  $ with the same space between points $ \lambda$, where $ \gamma \left( \lambda , x _{0} \right) = \left\lbrace n \lambda + x_{0} \vert n \in\mathbb{Z} \right\rbrace  $, and $ x _{0} \in \left[ 0 , \lambda \right)  $. This way of choosing $ x _{0} $ fixes the superselected sector, restricting the dynamics to a lattice $ \gamma \left( \lambda , x _{0} \right)   $ and work on separable Hilbert space $ \textit{H} ^{x_{0}} _{poly} $ consisting of wave functions which are non-zero only on the lattice.

Hence, the Schr\"{o}dinger equation and the associated eigenvalue problem becomes a difference equation for the wave function in coordinate representation
\begin{equation}
\psi _{\mu + 1} + \psi _{\mu - 1} = 2 \left\lbrace  1 -\frac{ m \lambda ^{2}}{\hbar ^{2}} \left[ E - V \left( x \right)  \right] \right\rbrace  \psi _{\mu}.  
\end{equation}

In contrast, in the momentum representation, it is generically a differential equation for $ 
\varphi \left( p \right) $ :
\begin{equation}
\frac{\hbar ^{2}}{m \lambda ^{2} } \left[  1- \cos \left( \frac{p \lambda}{\hbar} \right) \right] \varphi \left( p \right) = \left[ E - V \left( -i \hbar \partial _{p} \right) \right] \varphi \left( p \right) . 
\end{equation}

Working on $ \gamma \left( \lambda , x _{0} \right) $ restricts momentum wave functions $ \varphi \left( p \right) $ to periodic functions of period $ \frac{2 \pi \hbar}{\lambda} $ with the inner product formula (\ref{inner}) reducing to:
\begin{equation}
\left\langle x_{\mu} \vert x_{\nu} \right\rangle = \left\langle x_{\mu } \right\vert \Bigg( 
\frac{\lambda}{2 \pi  \hbar} \int _{-\frac{\pi \hbar}{\lambda }} ^{\frac{\pi \hbar}
{\lambda }} dp \vert p  \left\rangle  \right\langle  p \vert \Bigg) \left\vert 
x_{\nu}\right\rangle = \delta _{\mu \nu} , \label{inner2}
\end{equation}
and $ p \in \left( -\frac{\pi \hbar}{\lambda } , \frac{\pi \hbar}{\lambda } \right)$. Note that the identity operator (readed from (\ref{inner2})) on such subspace serves to define the inner product on $ \textit{H} ^{x _{0}} _{poly} $ in the momentum representation.

The dynamical quantum evolution of physical states can be described by the polymer propagator \cite{Flores}, which can be defined in the usual way. For time-independent Hamiltonian this is
\begin{equation}
k _{\lambda } \left( x _{\mu} , t ; x _{\nu} , t_{0} \right) = \left\langle x_{\mu } 
\right\vert e^{-i \frac{\widehat{H} _{\lambda} \left( t-t_{0} \right)}{\hbar}} \left\vert x_{\nu 
}\right\rangle , \label{polymer propagator}
\end{equation}
where we have chosen $ x _{0} = 0 $, so that $ x _{\mu} = \mu \lambda$. Hence, given an initial physical state at $ t = t _{0} $ , i.e. $ \left\vert x_{\nu } , t _{0} \right\rangle $, the state of the system for latter times in coordinate representation is given by
\begin{equation}
\psi \left( x _{\mu} , t \right) = \sum _{\nu} k _{\lambda} \left( x _{\mu} , t ; x _{\nu} 
, t_{0} \right) \psi \left( x _{\nu} , t _{0} \right).
\end{equation}
From its definition, it follows that the polymer propagator satisfy the standard consistency requirements to implement well-defined quantum evolution.

We conclude this brief review by pointing out that the polymer dynamics is equivalent to the conventional discrete approximation to the Schr\"{o}dinger equation when working on an superselected sector, but the conceptual difference is that in the polymer theory the lattice spacing is a fundamental constant of the theory.

\section*{Acknowledgements}
A. M. would like to thanks L. F. Urrutia and A. Frank for useful discussions, comments and suggestions.


\begin{thebibliography}{99}

\bibitem{Hossain2} Hossain G. M., Husain V. and Seahra S. S. 2010 Phys. Rev. D \textbf{81} 024005

\bibitem{Chacon} Chac\'{o}n-Acosta G., Manrique E., Dagdug L. and Morales-T\'{e}cotl H. A. 2011 SIGMA \textbf{7} 100

\bibitem{DIT} Mart\'{i}n-Ruiz A. 2014 Diffraction in Time of Polymer Particles arXiv:1406.3393

\bibitem{Eisberg} Eisberg R. and Resnick R. 1985 \textit{Quantum Physics of Atoms, Molecules, Solids, Nuclei, and Particles} 2nd ed. (John Wiley and Sons)

\bibitem{wigner} Salecker H. and Wigner E. P. 1958 Phys. Rev. \textbf{109} 571

\bibitem{davies} Davies P. C. W. 2005 Am. J. Phys. \textbf{73} 23

\bibitem{olkhovsky} Olkhovsky V. S. and Recami E. 1992 Physics Reports \textbf{214} 339

\bibitem{razavy} M. Razavy, \textit{Quantum Theory of Tunneling} (World Scientific, 2003) 

\bibitem{Gradshtein} Gradshteyn I. S. and Ryzhik I. M. 1994 \textit{Table of Integrals, Series, and Products}, Edited by A. Jeffrey and D. Zwil- linger 4th Edition (Academic Press, New York)

\bibitem{gorji} Gorji M. A., Nozari K. and Vakili B. 2014 Physics Letters B \textbf{735} 62

\bibitem{Hossain} Hossain G. M., Husain V. and Seahra S. S. 2010 Phys. Rev. D \textbf{82} 124032

\bibitem{Seahra} Seahra S. S., Brown I. A., Hossain G. M. and Husain V. 2012 Journal of Cosmology and Astroparticle Physics \textbf{2012} 041

\bibitem{Corichi} Corichi A., Vukasinak T. and Zapata J. A. 2007 Phys. Rev. D \textbf{76} 044016

\bibitem{Martin} Mart\'{i}n-Ruiz A., Bernal J. and Carbajal-Dom\'{i}nguez A. 2014 Journal of Modern Physics \textbf{5} 44

\bibitem{Ashtekar} Ashtekar A., Fairhurst S. and Willis J. L. 2003 Classical and Quantum Gravity \textbf{20} 1031

\bibitem{Flores} Flores-Gonz\'{a}lez E., Morales-T\'{e}cotl H. A. and Reyes J. D. 2013 Annals of Physics \textbf{336} 394

\end{thebibliography}
\end{document}